\documentclass[12pt]{article} 

\begin{document} 
\topmargin 0pt 
\oddsidemargin 0mm
\renewcommand{\thefootnote}{\fnsymbol{footnote}}
\begin{titlepage}
\vspace{5mm}

\begin{center}
{\Large \bf Is there a Gravitational Thomas Precession ?} 
\vspace{6mm}

{\large Harihar Behera$^a$\footnote{email: harihar@iopb.res.in} }\\
\vspace{5mm}
{\em
$^{a}$Patapur, P.O.-Endal, Jajpur-755023, Orissa, India\\}
\vspace{3mm}
\end{center}
\vspace{5mm}
\centerline{{\bf {Abstract}}}
\vspace{5mm}
Gravitational Thomas Precession ( GTP ) is the name given to Thomas
Precession when the acceleration is caused by a gravitational force
field. The contribution of the GTP to the the anomalous perihelion advance
of the  orbit of Mercury  is here estimated at
$\dot{\tilde{\bf\omega}}_{GTP}\,=\,21\cdot49\,\left[\,1\,+\,\frac{(\,\vec
    L\,\cdot\,\vec S\,)}{L^{2}}\,\right]
\mbox{arcsec/century}\,$,where $\,\vec L\,$ and $\,\vec S\,$
respectively represents the orbital angular momentum and the spin
angular momentum of Mercury .This effect seems to be of some  serious
concern for the General Relativity. \\


\end{titlepage}
\section{Introduction}
The Thomas precession\cite{1,2,3} is purely kinematical in origin
\cite{2}. If a component of acceleration $ (\vec a)$ exists
perpendicular to the velocity $ \vec v $, for whatever reason, then
there is a Thomas Precession, independent of other effects
\cite{2}. When the acceleration is caused by a gravitational force
field, the corresponding Thomas Precession is reasonably referred to
as the Gravitational Thomas Precession (GTP). Given the physics
involved in the Thomas Precession, the possibility of the existence of 
the GTP in planetary motion can not be ruled out in
principle. However, the very existence of the GTP in planetary motion
and its contribution to the perihelion advance of a planet as shown in 
this letter seems to be of some concern for the standard general
relativistic explanation for the observed anomalous perihelion advance 
of Mercury \cite{4,5,6}.        \\

\section{ The GTP's contribution to the Perihelion Advance  }     
The Thomas Precession frequency $ \vec\omega_{T} $ in the
non-relativistic
 limit (i.e., when $ v << c $) is given by \cite{2,3} 
\begin{equation}
\vec\omega_{T}\,=\,\frac{1}{2c^{2}}(\vec a \times \vec v ), 
\end{equation} 
where the symbols have there usual meanings. For a planet (say Mercury)
 moving around the Sun, the acceleration $ \vec a $ is predominately
 caused 
 by the Newtonian gravitational field of the Sun,viz., 
\begin{equation} 
\vec a\,=\,-\,\frac{GM_{\odot}}{r^{3}}\,\vec r\,\,, 
\end{equation} 
where the symbols have their usual meanings. Thus, from Eqs.$(1)$ and
$(2)$ we
 get the GTP frequency of the planet in question as 
\begin{equation} 
\vec \omega_{gT}\,=\,-\,\frac{GM_{\odot}}{2c^{2}r^{3}}\,(\vec r \times
\vec v )\,\,, 
\end{equation} 
where $ \vec v $ is the velocity of the planet.Since the angular
momentum of the planet is $\,\vec L\,=\,m(\vec r\times\vec v)\,$,
Eq.$(3)$ can be re-written as
\begin{equation} 
\vec \omega_{gT}\,=\,-\,\frac{GM_{\odot}}{2mc^{2}r^{3}}\,\vec L\,\,, 
\end{equation}
If, as Thomas first pointed out, that coordinate system rotates, then
the total time rate of change of the  angular momentum $\,\vec J\,$\footnote{For Thomas $\vec J\,=\,\vec S\,$, the spin angular momentum; but here we consider a more general term $\vec J\,=\,\vec L\,+\,\vec S\,$,$\vec
  J\,$ representing the total ( orbital\,+\,spin\,) angular momentum of the particle under 
  consideration. } or more
generally, any vector $\,\vec A\,$ is given by the well known result
\cite{2,3},
\begin{equation} \left(\frac{d\vec A}{dt}\right)_{\mbox{nonrot}}\,\,\,=\,\left (\frac{d\vec A}{dt}\right)_{\mbox{rest\,\,frame}}\,\,\,\,+\,\,\vec\omega_{T}\,\times\,\vec A
\end{equation}
where $\,\vec\omega_{T}\,$ is the angular velocity of rotation found
by Thomas. When applied to the total angular momentum $\,\vec J\,$,
Eq.$(5)$ gives an equation of motion:
\begin{equation} \left(\frac{d\vec J}{dt}\right)_{\mbox{nonrot}}\,\,\,=\,\left(\frac{d\vec J}{dt}\right)_{\mbox{rest\,\,frame}}\,\,\,\,+\,\,\vec\omega_{T}\,\times\,\vec J
\end{equation}
The corresponding energy of interaction is
\begin{equation} U\,=\,U_{0}\,+\,\vec J\,\cdot\,\vec\omega_{T}\,=\,U_{0}\,+\,\vec L\,\cdot\,\vec\omega_{T}\,+\,\vec S\,\cdot \vec\omega_{T}
\end{equation}
where $\,U_{0}\,$ is the energy corresponding to the coupling of
$\,\vec J\,$ to the external fields - say the Coulomb field in atomic
case ,nuclear field in nuclear case and the Newtonian gravitational
field in the planetary case. The origin of the Thomas precessional
frequency $\,\vec\omega_{T}\,$ is the acceleration experienced by the
particle as it moves under the action of external forces\cite{2}.Since 
the nature of the external forces is not specified, the result
obtained in Eq.(7) is valid for all type of force fields which cause
accelerations of whatever nature. When applied to the gravitodynamic
problems in solar system where the acceleration of a planet is caused
by a Newtonian force field Eq.(7) takes the form
\begin{equation} U_{g}\,=\,U_{0g}\,+\,\vec J\,\cdot\,\vec\omega_{T}\,=\,U_{0g}\,+\,\vec L\,\cdot\,\vec\omega_{gT}\,+\,\vec S\,\cdot \vec\omega_{gT}
\end{equation}
where $\,U_{0g}\,$ is the Newtonian potential energy of the planet
under consideration and $\,\vec\omega_{gT}\,$ is given by Eq.(4).We
then have
\begin{equation} U_{g}\,=\,-\,\frac{k}{r^{2}}\,-\,\frac{h_{1}}{r^{3}}\,-\,\frac{h_{2}}{r^{3}} 
\end{equation}
where $\,k\,=\,GM_{\odot}m\,$ and  
\begin{equation}h_{1}\,=\,\frac{GM_\odot L^{2}}{2mc^{2}} 
\end{equation}
\begin{equation}h_{2}\,=\,\frac{GM_\odot}{2mc^{2}}(\vec L\cdot\vec S). 
\end{equation}
Thus we see the gravitational Thomas precession introduced two
potentials of the form $\,1/r^{3}\,$ into the Kepler
problem.What effect will result from the introduction of these
potentials ? It is  shown in \cite{3} that if a potential with
$\,1/r^{3}\,$  form
is added to a central force perturbation of the bound Kepler problem,
the orbit in the bound problem is an ellipse in a rotating coordinate
system. In effect the ellipse rotates, and the periapsis appears to
precess.If the perturbation Hamiltonian is
\begin{equation}\bigtriangleup H\,=\,-\,\frac{h}{r^{3}},\,\,\,\,\,\,\,\,\,(\mbox{\,h\,=\,some constant\,})\,
\end{equation}
then it predicts \cite{3} a precession of the perihelion of a planet arising
out of the perturbation Hamiltonian (of the form as in Eq.(12)) at an
average  rate of
\begin{equation} 
\dot{\tilde{\bf\omega}}\,=\,\frac{6 \pi\,k\,m^{2}h }{\,\tau\,L^{4}}
\end{equation}
where $\,k\,=\,GM_{\odot} m $ and $\,\tau\,$ is the classical period of
revolution of the planet around the sun. It is to be noted that the so-called Schwarzschild spherically symmetric solution of the Einstein field equations corresponds to an additional Hamiltonian in the Kepler problem \cite{3,7} of the form of Eq.(12) with
\begin{equation} h\,=\,h_{E}\,=\,\frac{GM_{\odot}L^{2}}{mc^{2}}
\end{equation}
so that Eq.(13) becomes 
\begin{equation}
\dot{\tilde{\bf\omega}}_{E}\,=\,\frac{6\,\pi\,k^{2}\,}{\,\tau\,L^{2}\,c^{2}}\,=\, \frac{6\,\pi\,GM_{\odot}}{\,\tau\,c^{2}\,a\,(\,1\,-\,e^{2}\,)}
\end{equation}
where we have used the relation
$\,L^{2}\,=\,GM_{\odot}\,m^{2}\,a\,(\,1\,-\,e^{2})\,$. Eq.(15)
represents Einstein's expression for the anomalous perihelion advance
of a planet's orbit. Likewise the contributions to the perihelion
advance arising out of the Thomas precession can be estimated by
taking the $\,h\,$ in Eq.(13) as
\begin{equation}\,h\,=\,h_{1}\,+\,h_{2}
\end{equation}
where $\,h_{1}\,$ and $\,h_{2}\,$are respectively given by Eq.(10) and
Eq.(11).Then the GTP contribution comes out as
\begin{equation}
\dot{\tilde{\bf\omega}}_{GTP}\,=\,\frac{1}{2}\,\dot{\tilde{\bf\omega}}_{E}\,\left[\,1\,+\,\frac{(\,\vec L\,\cdot\,\vec S\,)}{L^{2}}\,\right]
\end{equation}
For Mercury, the value of
$\,\dot{\tilde{\bf\omega}}_{E}\,=\,42\cdot98\,$ arcsec/century - a
well known data \cite{3,4,5,6}.Hence the GTP's contribution to the
perihelion advance of Mercury's orbit is predicted at
\begin{equation}
\dot{\tilde{\bf\omega}}_{GTP}\,=\,21\cdot49\,\left[\,1\,+\,\frac{(\,\vec
    L\,\cdot\,\vec S\,)}{L^{2}}\,\right] \mbox{arcsec/century}
\end{equation}
This is in no way an insignificant contribution and should therefore be taken
into account in any relativistic explanation for the anomalous
perihelion advance of Mercury if it is not there. \\
\section{Concluding Remarks}
The Modern observational value of the anomalous perihelion advance of
Mercury is at $ \delta\dot{\tilde{\bf\omega}}\,\approx\,43'' $ per
century \cite{6}. For Mercury, General Relativity predicts this
phenomenon at                                                                  \begin{equation} 
\delta \dot{\tilde{\bf\omega}}_{GR}\,=\,\left[
  42\cdot98''\,+\,1\cdot289''(\,J_{2\odot}/{10}^{-5}\,)\right]\,\,{\rm{per\,\,century},}
\end{equation}
where $\,J_{2\odot}$ is the magnitude of solar quadrupole moment of the 
sun - a definite  value of which has still to be determined \cite{4}. Although
the Solar
quadrupole moment contribution is a Newtonian contribution, it was
taken account of because it was not there in the standard
expression. What about the GTP's contribution ? Is it already there in
the standard explanation ? The answer seems to be a negative one, because
the GTP's contribution contains a spin-orbit coupling factor ( see
Eq.(18) ),which is absent in Einstein's formula,viz.Eq.(15). Further
the Thomas precession and the Schwarzschild solution represent two
unrelated phenomena as either can exist independent of the other. The
Schwarzschild effect is a curved space-time phenomenon while Thomas
precession is a flat space-time phenomenon. So from these points of
view it seems the GTP contribution is not inherent in the standard
formula. So, if the existence of the GTP as an independent entity  can
not be denied in principle, it should be included in the general
relativistic formula.But such  an inclusion will have serious
implications for the general relativity's success in explaining the
observed perihelion precession of Mercury in particular and other
planets in general. \\
\textbf{Acknowledgments}\\   
 The author acknowledges the help received from Institute of Physics,
 Bhubaneswar for using its Library and Computer Centre for this
 work.\\ 
 
\end{document}